\documentclass[a4paper,12pt]{article} 
\usepackage[T1]{fontenc}
\usepackage[utf8]{inputenc}
\pdfoutput=1
\usepackage{amsmath}
\usepackage{amsfonts}
\usepackage{amssymb}
\usepackage[english]{babel}
\usepackage[pdftex]{graphicx,color} 
\usepackage[mathscr]{eucal}

\begin{document}
\title{Coloured loops in $4D$ and their effective field representation}
\author{~L.~E.~Oxman, ~G.~C.~Santos-Rosa, B.~F.~I.~Teixeira\\ \\
Instituto de F\'{\i}sica, Universidade Federal Fluminense,\\
Campus da Praia Vermelha, Niter\'oi, 24210-340, RJ, Brazil.}
\date{\today}
\maketitle

\begin{abstract} 
 
Gaining insight about ensembles of magnetic configurations, that could originate the confining string tension between quarks, constitutes a major concern in current lattice investigations. This interest also applies to a different approach, where gauge models with spontaneous symmetry breaking are constructed to describe the confining string as a smooth vortex solution. In this article, we initially show how to incorporate non Abelian information into an ensemble of monopoles in $4D$, characterized by phenomenological parameters. Next, using some recent techniques developed for polymers, we were able to relate the coloured ensemble with a non Abelian gauge model. This could offer an interesting perspective to discuss some of the different approaches to describe confinement.

\end{abstract}

\section{Introduction}

The difficulties posed by quark and gluon confinement are so deep that the mechanism underlying this phenomenon is still not understood. This is in
contrast to what happens at high energies where the strong interactions display
asymptotic freedom \cite{ref1}, and perturbative calculations in Yang-Mills (YM)
theories permit to make contact with observations. The main problem is that while  
confinement is associated with dimensionful scales, such as the confining string tension and width, the pure Yang-Mills gluon theory, at the classical level, only contains a dimensionless coupling constant. Then, confinement should be explained as a dimensional transmutation, driven by nonperturbative quantum effects.   

Some interesting scenarios for confinement have been proposed. For example, according to the dual superconductivity scenario \cite{N}-\cite{3}, the condensation of magnetic objects could generate an electric flux tube that confines cromoelectric charges. These ideas have been approached by different communities. In particular, they have been explored in the lattice, which provides a natural nonperturbative tool to compute Wilson loop averages and extract the interquark potential. In the last years, various ensembles of magnetic defects that could capture the main contribution to the sum over lattice configurations have been analyzed (for a review, see ref. \cite{ref3}). 

In the continuum, similar ideas are hard to explore, as a nonperturbative definition of the path-integral measure is lacking. It could be that magnetic objects appear as defects when fixing a gauge \cite{ref3},
or maybe they show up as semiclassical configurations to be path-integrated \cite{Diakonov}, \cite{Faber}. A different theoretical approach to dual superconductivity is based on constructing gauge models with spontaneous symmetry breaking (SSB), that could describe the confining string \cite{Bali} as a smooth classical vortex solution  \cite{Baker}-\cite{notes}. 
Although these models cannot be derived from first principles, they could provide an understanding, in the same sense the Ginzburg-Landau model represents an understanding of BCS superconductors, which is completed by the underlying  Cooper pair condensation mechanism. 

Similarly to lattice studies, a guiding principle to prefer a given dual superconductor model to another would be if it better reproduces the properties of the confining string obtained from Monte Carlo simulations, for different gauge groups and quark representations \cite{GSY,DPRV}. Now, it is reasonable expecting that the descriptions in terms of magnetic ensembles and SSB gauge models that best fit the simulations should be somehow related. Although these best descriptions have not yet been established, gaining information about the relation between a given ensemble and an effective SSB gauge model could offer an interesting perspective on the problem of confinement. This is the point of view that we will  follow in this article.\vspace{.3cm}  
 
Having introduced the general setting, let us now discuss some important results obtained by the different communities. In the lattice, several studies tell us that monopoles (loops in $4D$), originated as defects in maximally Abelian gauges, could be important to reproduce the string tension between a quark antiquark pair in the fundamental representation \cite{ref4}-\cite{ref6}. The associated SSB gauge models are Abelian \cite{Baker, Baker1}, \cite{antonov},  with the Higgs field effectively representing the loop ensemble, and  the vortex nicely adjusting the potential as a function of the $q\bar{q}$ separation \cite{Baker,Baker1}.
However, Abelian scenarios cannot explain the asymptotic dependence of the lattice potential  on the way $Z(N)$, the center of 
$SU(N)$, is realized in a given quark representation, see ref. \cite{ref3}. In the lattice  community, this led to the idea of center dominance, where center vortex configurations, originated as defects in center gauges, are capable of explaining $N$-ality at asymptotic distances \cite{ref7}-\cite{ref11}. Configurations containing correlated monopoles and center vortices could also be very promising to describe the different properties of the interquark potential \cite{ref12}-\cite{ref19}. Note that unlike ensembles of monopole loops, center vortices are worldsurfaces in $4D$, so they cannot be naturally associated with an effective {\it field} description  \cite{ref11}. 

On the analytical side, motivated by the pioneering work on supersymmetric non Abelian monopoles and strings in refs. \cite{HD}-\cite{SY}, non supersymmetric gauge models with vortex solutions that represent a confining string with the right $N$-ality properties were proposed and analyzed. They correspond to dual superconductors supporting $Z(N)$-vortices \cite{Konishi-Spanu} and non Abelian $k$-strings \cite{GSY}. The former were constructed in terms of a set of adjoint Higgs fields, that drive an $SU(N) \to Z(N)$ SSB transition. It is important to underline that these confining $Z(N)$ center vortices are {\it smooth} solutions to the dual superconductor equations \cite{Konishi-Spanu}, \cite{deVega}-\cite{HV}, and  should not to be confused with the center vortex magnetic configurations discussed in lattice $SU(N)$ pure Yang-Mills. In addition, besides the confining potential between a quark/antiquark colourless pair, we would also like to describe a confining potential between a colour nonsinglet pair. This possibility is not ruled out by QCD, as the confining string could be in an excited nonsinglet colour state, that couples to the quarks to form an overall colourless object. These $qg\bar{q}'$ hybrid mesons, where $g$ represents a valence gluon, have been observed in the lattice and current experiments are devoted to detect them \cite{hybrids}-\cite{DuE}. They have also been recently incorporated in SSB generalized gauge models with a richer adjoint Higgs field structure \cite{ref28}.

The purpose of this article is initiating a theoretical study aiming at determining what is the ensemble underlying non Abelian superconductor models with a given Higgs field content. The picture that will be delineated is that ensembles of coloured monopoles could be important to reproduce some of the main features of these models.
In particular, monopoles characterized by weights of the adjoint representation will be naturally related with non Abelian models with adjoint Higgs fields, which possesses good $N$-ality properties. Then, in the light of the above discussion, although Abelian monopole lattice configurations cannot explain $N$-ality, maybe non Abelian lattice monopoles could provide an alternative/complementary picture to lattice center vortices as sources of $N$-ality. Non Abelian monopoles and non Abelian electric-magnetic dualities have been conjectured in ref. \cite{GNO}, while the relevance of such configurations has been explored in the lattice \cite{Kondo-col}. 

In this article, using some recent techniques developed for polymers, we relate  an ensemble of coloured loops in $4D$ with a large distance effective non Abelian SSB gauge model. In this respect, it is important to emphasize that ``effective'' refers to the large distance representation of the ensemble which, similarly to the SSB model, is characterized by phenomenological dimensionful parameters from the beginning.

The article is organized as follows. In section 2 we discuss ensembles of
loops in $4D$ and show how to include non Abelian information. In \S 3, we discuss the appropriate path-integral discretization and measure for the propagation of colour. In \S 4, we introduce the natural phenomenological properties characterizing one-dimensional objects, depending on their length (tension) and on their curvature (stiffness). 
In \S 5, we analyze the coloured Chapman-Kolmogorov equation that generates the end-to-end probabilities. In \S 6, we derive the diffusion equation for a one-dimensional interacting non Abelian object, and perform the ensemble integration to obtain the effective non Abelian field description. Finally, in section 7, we present our conclusions. 

\section{$SU(N)\to Z(N)$ dual superconductor models}

The aim of this article is analyzing the relation between the phenomenological
properties that could characterize
ensembles of one-dimensional objects and the associated effective descriptions.
This could prove useful as a complement to the lattice, and a guide to discuss
models that capture the YM infrared behavior.
Some theoretical steps have already been taken in this direction. Mon\-o\-poles,
center vortices, as well as chains formed by them, 
can be accommodated as topological defects of different local bases in colour
space 
\cite{ref20}-\cite{ref22}. Recently, we showed that ensembles of chains in $3D$
generate the effective model,
\begin{equation}
\overline{D_\mu V} D_\mu V + m^2 \, \bar{V} V + \alpha \, (\bar{V} V)^2 
+ \beta \, (V^N + \bar{V}^N) + \frac{\gamma}{2} \, \lambda_\mu \lambda_\mu +
\frac{\delta}{2} \, F_{\mu \nu}F_{\mu \nu} \;,
\label{modelg}
\end{equation}
\begin{equation}
D_\mu V = [\partial_\mu + i\, (2\pi/g)\, \lambda_\mu] V 
\makebox[.5in]{,}
F_{\mu \nu} = \partial_\mu \lambda_\nu - \partial_\nu \lambda_\mu
\;,
\end{equation}
where $V$ represents the vortex sector and the dual field $\lambda_\mu$
represents the off-diagonal sector of YM theories \cite{ref23}. Thus, when the
massive $\lambda_\mu$ is decoupled, we made contact with the well-known vortex 
model proposed by t'Hooft to describe confinement in $3D$ $SU(N)$ YM theories, 
\begin{equation}
\partial_\mu \bar{V} \partial_\mu V + m^2 \, \bar{V} V + \alpha \, (\bar{V}
V)^2 
+ \beta \, (V^N + \bar{V}^N) \;,
\end{equation}
which displays a global {\it magnetic} $Z(N)$ symmetry. When the vortex is an
elementary
excitation ($m^2 > 0$), there is no SSB. If vortices condense, SSB occurs
($m^2 < 0$) 
and the formation of a domain wall between a heavy quark-antiquark pair
leads to an area law for the Wilson loop \cite{3}.

In ref. \cite{ref23}, to obtain the $V$-sector in eq. \eqref{modelg}, the
end-to-end probability for a single interacting vortex, containing some
phenomenological parameters, was an essential ingredient. In that reference,
following  techniques developed for polymers \cite{ref24}, we determined the
weight for a one-dimensional object in $3D$, interacting with a scalar field
$\phi$ (describing excluded volume effects), and the dual vector field
$\lambda_\mu$. Combining this elementary block by joining the center
vortex endpoints in groups of $N$, virtual process where $N$ center vortices are
created at a monopole-like instanton were incorporated, thus originating the
different terms for the effective vortex field.

Obtaining an effective model for vortex-monopole chains in $4D$ is a hard
problem as the center vortex component is given by  two-dimensional surfaces,
while effective field theories are only naturally associated with
one-dimensional objects. On the other hand, monopoles are looplike in
$4D$ so that polymer techniques can be followed again \cite{ref25}.  In addition, we have recently proposed a Yang-Mills-Higgs (YMH) $4D$ effective model \cite{ref28},
\begin{equation}
\frac{1}{2} \langle D_\mu \psi_I , D^\mu \psi_I\rangle + V_{\rm Higgs}(\psi_I) +
\frac{1}{4} \langle F_{\mu \nu}, F^{\mu \nu}\rangle \;,
\label{modelg4}
\end{equation}
\begin{equation}
D_\mu=\partial_\mu-ig [\Lambda_\mu, ~]\makebox[.5in]{,}
F_{\mu \nu}= \partial_\mu \Lambda_\nu -\partial_\nu \Lambda_\mu -i g
[\Lambda_\mu,  \Lambda_\nu]  \;,
\label{somed}
\end{equation}
with gauge group $SU(N)$, containing a set of adjoint Higgs fields $\psi_I \in
\mathfrak{su}(N)$, where $V_{\rm Higgs}(\psi_I)$ involves the natural invariant terms formed with the metric and pairs of Lie algebra elements\footnote{The symbol $\langle ,\rangle $ stands for the Lie algebra metric.}. Up to quartic order, these terms are,
\begin{equation}
\langle \psi_I,\psi_J \rangle
\makebox[.3in]{,} \langle\psi_I,\psi_J \wedge \psi_K\rangle
\makebox[.3in]{,}  \langle \psi_I\wedge \psi_J,\psi_K \wedge \psi_L\rangle
\makebox[.3in]{,}
\langle \psi_I,\psi_J \rangle \langle \psi_K,\psi_L \rangle
\label{terms}
\;,
\end{equation}
where $\wedge $ stands for the Lie algebra product\footnote{the $i$ factor turns
the product of hermitian fields closed.},
\begin{equation}
\psi_I \wedge \psi_J = -i [\psi_I,\psi_J]\;.
\end{equation}
Although the minimum number of fields required to completely break $SU(N)$ down
to $Z(N)$ is $N$ \cite{deVega}-\cite{HV}, in ref. \cite{ref28} we considered a larger number,  $\psi_A$, $A=1,\dots,
N^2-1$ (here, $A$ is a flavour index), to write a flavour symmetric Higgs potential with global group $Ad(SU(N))$,
\begin{eqnarray} 
 V_{\rm Higgs}= c+ \frac{m^2}{2}\, \langle \psi_A ,\psi_A \rangle
+\frac{\gamma}{3} \,f_{ABC} \langle \psi_A \wedge \psi_B,\psi_C \rangle
+\frac{\lambda}{4}\, \langle \psi_A \wedge \psi_B,\psi_A\wedge \psi_B \rangle 
\;.
\label{Vf}
\end{eqnarray}
What makes this type of models interesting is that, in the phase where the gauge
symmetry is spontaneously broken, they can describe not only the confining
string between a $q\bar{q}$ pair of (red/anti-red, green/anti-green, \dots)
external quarks, but also other possible excited states. In particular,
$qg\bar{q}'$ hybrid mesons \cite{hybrids}-\cite{DuE}, formed by say a
red/anti-green pair of external quarks bound by an anti-red/green valence gluon,
can be written as a $\psi_A$, $\Lambda_\mu$ non Abelian configuration. While the normal string is a center vortex of the effective
model \eqref{Vf}, the excited string is formed by a pair of center vortices interpolated by
a monopole, which is identified with the valence gluon. 

Similarly to the understanding we obtained for the $3D$ vortex model \eqref{modelg}, in terms of an underlying ensemble of chains, here we would like to give the initial
steps to determine the ensemble 
underlying the Higgs sector for the type of models in eq. \eqref{modelg4}. For this
reason, we are interested in introducing non Abelian information
into the characterization of an ensemble of interacting loops in $4D$, and
compute their effective field description. For a discussion of monopole ensembles in an Abelian context,
see ref. \cite{ref27}.

\section{Ensembles of coloured loops}

Let us consider a $\mathscr{D}$-dimensional matrix representation of the Lie algebra $\mathfrak{su}(N)$. The generators $T_A$, $A=1,\dots, N^2-1$, with matrix elements 
$T^{ab}_A$, $a,b =1,\dots, \mathscr{D}$, satisfy,
\begin{equation}
[T_A,T_B]=if_{ABC} T_{C}
\makebox[.5in]{,}
\langle T_A,T_B\rangle = \delta_{AB}\;.
\label{Lie-alg}
\end{equation} 
In particular, $\mathscr{D}=N$ for the fundamental representation, and  $\mathscr{D}=N^2-1$ for the adjoint. 
As usual, the metric is defined as,
\begin{equation}
\langle X ,Y\rangle =Tr \left(Ad(X)^\dagger Ad(Y)\right),
\label{metric}
\end{equation}
where $Ad(X)$ is a linear map of $X\in \mathfrak{su}(N)$ into the adjoint
representation generated by the hermitian matrices $M_A$, with elements $M_A|_{BC}=-i f_{ABC}$, satisfying,
\begin{equation}
\left[M_{A},M_{B}\right]=if_{ABC} M_C\;,
\label{algeb}
\makebox[.5in]{,}
Tr(M_A M_B)=\delta_{AB}\;.
\end{equation}

According to ref. \cite{ref26}, the coupling of a particle worldline carrying a
non Abelian charge, in the $\mathscr{D}$-dimensional representation, is
implemented by the Lagrangian,
\begin{equation}
\mu \, (\dot{x}_{\mu}\dot{x}_{\mu})^{1/2} +\frac{1}{2} (\bar{z}_c \dot{z}_c -
\dot{\bar{z}}_c z_c) -ig\, 
\dot{x}_\mu I^A \Lambda_{\mu}^{A} \makebox[.5in]{,} I^{A}=T^{A}_{cd}\,
\bar{z}_{c} z_{d} \;,
\label{Bal}
\end{equation}
where $z_{c}$. $c=1,\dots,\mathscr{D}$, describes the colour degrees of freedom. 
Therefore, it is natural introducing the partition function for an ensemble of
coloured loops as,
\begin{equation}
Z = \sum_n\, \int [Dm]_n \, e^{-\left[ S^0 +S^{\rm int}\right]}\;,
\label{zetavm}
\end{equation}
where  $\sum_n$ sums over the number of loops, while $S^0$ and $S^{\rm int}$
are the free and interacting parts of the action, containing respectively the
terms, 
\begin{equation}
\mu \, (\dot{x}^{(k)}_{\mu}\dot{x}^{(k)}_{\mu})^{1/2}
\makebox[.5in]{,}
-ig  \dot{x}^{(k)}_\mu I^{A}\Lambda_{\mu}^{A}(x^{(k)})\;,
\end{equation}
where $k=1,\dots , n$ denotes each one of the loops in the $n$-sector. 

Besides the phenomenological parameter $\mu$, describing tensile properties, it will be fundamental to include in $S^0$ the effect of stiffness, measured by $1/\kappa$. Larger values of $\kappa$ correspond to more flexible objects. These are the simplest properties that can be associated with a worldline, that is, a weight that depends on its length and curvature. In Abelian ensembles, the presence of stiffness is supported by lattice calculations that show a strong correlation between different link 
orientations on monopole loops \cite{BPZ, BBPZ}.  

In addition, besides the non Abelian gauge field $\Lambda_{\mu}^{A}$, the interaction part
will also contain scalar and isovector loop interactions.
They can be reproduced by introducing fields 
$\phi(x)$, $\Phi^A(x)$, to be integrated with an appropriate weight $e^{-W}$.
For example, when $W$ is of the form,
\begin{equation}
W \propto -\int d^{4}x \, \phi^{2}\;,
\end{equation}
the path integral over $\phi$ would implement excluded volume effects
\cite{ref23}. Other possibilities for $W$ will be specified later.

Summarizing, the partition function in eq. \eqref{zetavm} shall be given by,
\begin{eqnarray}
\lefteqn{Z=\int [D\phi][D\Phi]\, e^{-W}\, \sum_n\, Z_n}, \nonumber \\
&& Z_n= \int [Dm]_n \, \exp \left[ {\sum_{k=1}^{n}\oint_{L_{k}} ds\;
\left(ig  \dot{x}^{(k)}_\mu I^{A}\Lambda_{\mu}^{A}(x^{(k)})-
I^{A}\Phi^{A}(x^{(k)})-\phi(x^{(k)})\right) - S^0}\right]\;,
\label{Zn}\nonumber \\
&& S^0=\sum_{k=1}^{n}{\oint_{L_{k}} ds\, \biggl[ \mu + \frac{1}{2} (\bar{z}_c
\dot{z}_c - \dot{\bar{z}}_c z_c)+ \frac{1}{2\kappa}\, \dot{u}^{(k)}_\mu
\dot{u}^{(k)}_\mu  \biggr]},
\end{eqnarray}
where $s$ is the arc length parameter and
${u}^{(k)}=\dot{x}^{(k)}$ is the unit tangent vector.

The measure $[Dm]_n$ must integrate over all possible $n$ closed monopole
worldlines, including the different lengths and shapes. That is, following refs.
\cite{ref23,ref27}, we define,
\begin{eqnarray}
\label{Dm}
[Dm]_n &\equiv& \frac{1}{n!}\int_{0}^{\infty}\;
\frac{dL_{1}}{L_{1}}\frac{dL_{2}}{L_{2}}...\frac{dL_{n}}{L_{n}} \;
\int_{\Re^{4}}\; d^{4}x_{1}d^{4}x_{2}...d^{4}x_{n} \nonumber \\ &\times& \int \;
[Dx^{(1)}(s)]_{(x_{1},L_{1})}[Dx^{(2)}(s)]_{(x_{2},L_{2})}...[Dx^{(n)}(s)]_{(x_{
n},L_{n})} \nonumber \\ &\times& \int \; \sum_{a_{1}}[Dz^{(1)}]_{(a_{1},a_{1})}
\; \sum_{a_{2}}[Dz^{(2)}]_{(a_{2},a_{2})}...\;
\sum_{a_{n},a_{n}}[Dz^{(n)}]_{(a_{n},a_{n})}\;.
\end{eqnarray}
As the monopoles are identical, a $1/n!$ factor arises. The measure,
\[
[Dx^{(k)}(s)]_{(x_{k},L_{k})}
\] 
represents  the integration over all closed curves with fixed length $L_k$, such
that at $s=0$ and $s=L_k$, it is verified $x(s)=x_k$. Noting that
the same closed curve can be generated with different $x_k$'s, a direct $d^4
x_k$ integration would lead to an overcounting, that scales as the length of the
curve. This is corrected by the $1/L_k$ factor \cite{ref27}. The measure
$[Dz^{(k)}]_{(a_{k},a_{k})}$ 
acts on the internal variables, and is designed to implement a propagation from an initial {\it definite} colour $a_k$, at $s=0$, to the same
final colour at $s=L_k$. Its detailed form shall be given in the next section.
These processes are to be summed over all possible colours $a_k=1,\dots,\mathscr{D}$ of
the representation of $SU(N)$ under consideration.

Now, we note in eq. \eqref{Zn}, with the help of eq. \eqref{Dm}, that the
partition function can be written as,
\begin{equation}
\label{zdefq}
Z =\int [D\phi][D\Phi]\, e^{-W}\,
e^{\int_{0}^{\infty}\frac{dL}{L}\; \int_{\Re^{4}} d^{4}x \,
\sum_{a} Q^{aa}(x,x,L)} \;,
\end{equation}
where we introduced the fundamental block,
\begin{eqnarray}
\lefteqn{Q^{ba}(x,x_0,L)= \int [Dx(s)]_{x_0,x,L}\;[Dz(s)]_{(a,b)}} 
\nonumber \\
&& \times\, e^{ -\int_{0}^{L} ds\, \left[\mu + \frac{1}{2} (\bar{z}_c \dot{z}_c
- \dot{\bar{z}}_c z_c)+ \frac{1}{2\kappa}\, \dot{u}_\mu \dot{u}_\mu 
+\phi(x) -ig\, \dot{x}_\mu \Lambda_{\mu}^{A}T^{A}_{cd}\, 
\bar{z}^{c} z^{d}+ \Phi^{A} T^{A}_{cd}\, 
\bar{z}^{c} z^{d}\right]} \;, 
\label{Qprin}
\end{eqnarray}
describing the end-to-end probability for a single open one-dimensional
object that propagates from location $x_0$ and colour $a$, at $s=0$, to $x$ and $b$, at $s=L$. The measure $[Dx(s)]_{x_0,x,L}$ represents the integral over
paths with fixed length $L$ such that, $x(0)=x_0$, $x(L)=x$. Then, it becomes
clear that further knowledge about the properties of the building block
$Q^{ba}(x,x_0,L)$ is essential to get an effective field representation of the
monopole ensemble. Note that the dimension of $Q^{ba}(x,x_0,L)$ is $L^{-4}$, as it is in fact obtained from the end-to-end probability density
$Q^{ba}(x,x_0,L)$ after setting $x=x_0$, where $Q^{ba}(x,x_0,L) \,d^4x$ is the
probability to have the endpoint of the curve on a volume $d^4x$, centered at
$x$, given that the initial point is at $x_0$. As a consequence, the exponent in
eq. \eqref{zdefq} is adimensional, as it should be. 

In the next section, we
review some fundamental properties of the coherent states for quantum systems
with non Abelian internal degrees of freedom.

\section{Coherent states} 

As an important step to deal with one-dimensional objects coupled to a non
Abelian field, we have to complete the description of the internal degrees of
freedom. In particular, how a definite colour is propagated in the dynamics
represented by \eqref{Bal}. 

Initially, let us consider the creation and annihilation operators
$\hat{a}_{c}^{\dagger}$ and $\hat{a}_{c}$, which act on the space of colour states, satisfying the commutation relations, 
\begin{eqnarray}
[\, \hat{a}_{c},\hat{a}_{d}^{\dagger}\,]=\delta_{cd}, \qquad
[\,\hat{a}_{c},\hat{a}_d\,]=[\,\hat{a}_{c}^{\dagger},\hat{a}_{d}^{\dagger}\,]=0 \;.
\end{eqnarray}
When applied on the vacuum, $\hat{a}_{c}^{\dagger}$ creates a state with
well-defined colour $c$. The corresponding occupation numbers are vanishing for every
colour but for the given colour $c$, where it takes the value one. In the space
of states, a basis   
can be defined in terms of the different occupation numbers, for every possible
colour, $|n_{1},\cdots,n_{\mathscr{D}}\rangle$. 

As usual, we have,
\begin{eqnarray}
|n_1,\dots, n_{\mathscr{D}}\rangle= \prod_{c=1}^{\mathscr{D}}\frac{1}{\sqrt{n_c
!}}(\hat{a}_{c}^{\dagger})^{n_c}|0,\dots ,0\rangle \;,
\end{eqnarray}
and we can define coherent states $|z \rangle =|z^1,\dots,z^{\mathscr{D}}\rangle $,
$z^{a}\in C$, with the properties,
\begin{eqnarray}
\hat{a}_{c}|z \rangle=z^{c}|z\rangle, \qquad
\hat{a}_{c}^{\dagger}|z\rangle=\frac{\partial}{\partial z^{c}}|z\rangle \;,
\end{eqnarray}
As a shorthand notation, we denote the set of variables and 
their conjugates as $z=(z^1,\dots,z^\mathscr{D})$ and $\bar{z}=(\bar{z}^1,\dots,
\bar{z}^\mathscr{D})$, respectively. 

The coherent states can be written as,
\begin{eqnarray}
| z\rangle=e^{z \cdot \hat{a}^{\dagger}}\, |0,\dots,0\rangle \;,
\end{eqnarray}
and their overlap is,
\begin{eqnarray}
\label{prodintcoe}
\langle z |z'\rangle=\langle 0|e^{\bar{z}\cdot \hat{a}}\, e^{ z' \cdot
\hat{a}^{\dagger}}|0\rangle=e^{\bar{z}\cdot z'}\;.
\end{eqnarray}
Any state $|\psi\rangle$ can be expanded in the occupation number basis,
\begin{eqnarray}
\lefteqn{|\psi\rangle=
\sum_{n_1,\dots,n_{d}}\frac{\psi_{n_1, \dots , n_{\mathscr{D}}}}{\sqrt{n_1
!}\dots\sqrt{n_{\mathscr{D}}!}}\,|n_1,\dots, n_{\mathscr{D}}\rangle } \nonumber \\
&& =\sum_{n_1,\dots,n_{\mathscr{D}}}\psi_{n_1, \dots , n_{\mathscr{D}}}
\prod_{c=1}^{\mathscr{D}}\frac{(\hat{a}_{c}^{\dagger})^{n_c}}{n_c !}|0,\dots
,0\rangle\;,
\end{eqnarray}
and its projection on the coherent state $|z\rangle$ is,
\begin{eqnarray}
\langle z|\psi\rangle &=&\sum_{n_1,\dots,n_{\mathscr{D}}} \psi_{n_1, \dots , n_{\mathscr{D}}}\,
\frac{(\bar{z}^{1})^{n_1}\dots (\bar{z}^{\mathscr{D}})^{n_{\mathscr{D}}} }{n_1 !\dots n_{\mathscr{D}} !} \;, 
\end{eqnarray}
that  is, $\langle z|\psi\rangle =\psi(\bar{z})$ is an anti-holomorphic function
of the variables $\bar{z}$. This will prove to be an important property to
construct a recurrence relation generating the sum over histories.

In the space of colour states, the identity operator is,
\begin{eqnarray}
\label{defelint}
\hat{I}=\int dz d\bar{z} \; e^{-\bar{z} \cdot z}\,|z\rangle\langle z|; \qquad
\qquad \int
dzd\bar{z}\equiv\int\prod_{c=1}^{\mathscr{D}}\Bigg[\frac{dz^{c}d\bar{z}^{c}}{2\pi
i}\Bigg],
\end{eqnarray}
namely,
\begin{eqnarray}
\label{deltacoe}
\psi(\bar{z})=\langle z|\hat{I}|\psi\rangle = \int dz' 
d\bar{z}'\; e^{(\bar{z}-\bar{z}')\cdot z'}\psi(\bar{z}') \;.
\end{eqnarray}

Let us now consider the weight $\langle f|e^{-\hat{H}L}|i\rangle$ for the
propagation of an initial state $|i\rangle$ to a final state $|f\rangle$. 

As usual, the interval $L$ can be divided into $M$ parts of length $\Delta L$, $L= M
\Delta L$, with $ M\to\infty$ and $\Delta L\to 0$, and insert $M$ identities
$\hat{I}$, to obtain,
\begin{eqnarray}
\langle z|e^{-\hat{H}L}|z_{0}\rangle=\int \prod_{i=1}^{M} dz_{i} 
d\bar{z}_{i}\, e^{-\bar{z}_{i}\cdot z_i}\, \langle z_{i+1}|e^{-\hat{H}\Delta
L}|z_{i} \rangle  \langle z_1 ,z_0\rangle \;,
\end{eqnarray}
where $z_{M+1}=z$. 

The natural interaction between the colour degrees of freedom, the non Abelian gauge field
and the isovector is,
\begin{equation}
\hat{H}=-ig\, u_\mu \Lambda_{\mu}^{A}\, T^{A}_{cd}\,
\hat{a}^{\dagger}_{c}\hat{a}_{d} + \Phi^{A}\, T^{A}_{cd}\,
\hat{a}^{\dagger}_{c}\hat{a}_{d}  \;.
\end{equation}
In this case, for small $\Delta L$, 
\begin{eqnarray}
\langle z_{i+1}|e^{-\hat{H}\Delta L}|z_{i}\rangle &\approx &\langle
z_{i+1}|z_{i}\rangle \, (1-{H}(\bar{z}_{i+1},z_{i})\Delta L) \nonumber \\
&\approx & e^{\bar{z}_{i+1}\cdot z_{i}}\, e^{-{H}(\bar{z}_{i+1},z_{i})\Delta
L}\;,
\end{eqnarray}
\begin{equation}
{H}(\bar{z},z') = -ig u_\mu \Lambda_{\mu}^{A}T^{A}_{cd}\, \bar{z}^{c} {z'}^{d} +
\Phi^{A}T^{A}_{cd}\, \bar{z}^{c} {z'}^{d}\;. 
\end{equation}
Therefore,
\begin{eqnarray}
\langle z|e^{-\hat{H}L}|z_{0}\rangle &=& \lim_{\Delta L\to 0}\int
\prod_{i=1}^{M} dz_{i} 
d\bar{z}_{i}\, e^{\sum_{j=1}^M \left[(\bar{z}_{j+1}-\bar{z}_j) \cdot
z_{j}-{H}(\bar{z}_{j+1},z_{j})\Delta L \right] }\, e^{\bar{z}_1 \cdot z_0}\;,
\nonumber \\
\label{cc}
\end{eqnarray}
For the transition between general initial and final states, 
\begin{eqnarray}
|\psi\rangle=\int dz_{0} d\bar{z}_{0}\; e^{-\bar{z}_{0}\cdot z_0}\, \psi
(\bar{z}_{0})\, |z_{0}\rangle \makebox[.5in]{,}
\langle \phi|=\int dz d\bar{z}\; e^{- \bar{z}\cdot z}\, \bar{\phi} (z)\, \langle
z| \;,
\end{eqnarray}
we can use,
\begin{eqnarray}
\lefteqn{\langle \phi|e^{-\hat{H}L}|\psi\rangle = \int dz d\bar{z}\;dz_{0}
d\bar{z}_{0}\;  e^{- \bar{z}\cdot z}\, e^{-\bar{z}_{0}\cdot z_0}\,
\bar{\phi}(z)\,\psi (\bar{z}_{0})\,\langle z|e^{-\hat{H}L}|z_{0}\rangle }
\nonumber \\
&&=\lim_{\Delta L\to 0}\int [Dz]_{(\psi,\phi)}\; e^{\sum_{j=0}^M 
\left[ \frac{1}{2}(\bar{z}_{j+1}-\bar{z}_j) \cdot z_{j}+\frac{1}{2}
\bar{z}_{j+1} \cdot (z_{j}-z_{j+1})\right] - \sum_{j=1}^M
{H}(\bar{z}_{j+1},z_{j})\Delta L  }\;, \nonumber \\
&& [Dz(s)]_{(\psi,\phi)} \equiv \prod_{i=0}^{M+1} dz_{i} 
d\bar{z}_{i} \;  e^{- (\bar{z}\cdot z)/2}\, e^{-(\bar{z}_{0}\cdot z_0)/2}\, 
\bar{\phi}(z)\, \psi(\bar{z}_{0}) \;,
\label{pcol}
\end{eqnarray}
thus making contact with the colour sector of Balachandran's Lagrangian in eq.
\eqref{Bal}, and a precise definition for the corresponding path-integral
measure. The limiting procedure in eq. \eqref{pcol} will be denoted as,
\begin{equation}
 \langle \phi|e^{-\hat{H}L}|\psi\rangle \equiv \int [Dz]_{(\psi,\phi)}\,
e^{-\int_0^L ds\,  {\cal L}_E}\;,
\label{pf}
\end{equation}
\begin{equation}
{\cal L}_E = \frac{1}{2} (\bar{z}_c \dot{z}_c - \dot{\bar{z}}_c z_c) -ig u_\mu
\Lambda_{\mu}^{A}T^{A}_{cd}\, 
\bar{z}^{c} z^{d} + \Phi^{A}T^{A}_{cd}\, 
\bar{z}^{c} z^{d}\;.
\end{equation}
In particular, for the transition between states with well-defined colours, $|\psi\rangle =| a\rangle$ and $|\phi\rangle =|b\rangle $,  the integration measure (to be used in eq. \eqref{Qprin}) is,
\begin{equation}
[Dz(s)]_{(a,b)} \equiv \frac{1}{{\cal M}}\, \prod_{i=0}^{M+1} dz_{i} 
d\bar{z}_{i} \;  e^{- (\bar{z}\cdot z)/2}\, e^{-(\bar{z}_{0}\cdot
z_0)/2}\,z^{b}\bar{z}_{0}^{a} \;,
\end{equation}
where the normalization factor ${\cal M}$ is such that, for $L=0$,
eq. \eqref{pf} gives $\delta_{ab}$.

\section{Chapman-Kolmogorov equation with colour} 

An important step to obtain the diffusion equation for one-dimensional objects
is to generate the end-to-end probability by means of a recurrence relation.
This is customary in the physics of polymers \cite{ref24}, and has been recently
used to derive the center vortex weight in $(2+1)D$, a fundamental block to
build the effective model in eq. (\ref{modelg}),
where no internal degrees of freedom were considered \cite{ref23}.

Now, in the colour sector, the discretized amplitude in eq. \eqref{cc} can be
generated by the recurrence relation,
\begin{equation}
\label{recinc}
q_{j}( \bar{z},z_0) = \int dz'd \bar{z}' \; e^{(\bar{z}-\bar{z}')\cdot z'}e^{-H(
\bar{z},z')\Delta L}\, q_{j-1}( \bar{z}',z_0)\;,
\end{equation}
with the initial condition,
\begin{equation}
q_{0}( \bar{z}_{1},z_0)=\langle z_{1}|z_0\rangle=e^{\bar{z}_{1}\cdot z_{0}} \;.
\end{equation}
Upon $M$ iterations, it is obtained,
\begin{equation}
 q_{M}( \bar{z}_{M+1},z_0) =
\langle z|e^{-\hat{H}L}|z_0\rangle \;.
\end{equation}

Therefore, in order to generate the full weight for a one-dimensional object
with colour, we shall consider the following Chapman-Kolmogorov equation in $n$
dimensions,
\begin{eqnarray}
\label{recesi}
q_{j}(x,x_0,u,u_0, \bar{z},z_0) = \int d^{n}x' d^{n-1}u'\, dz'd \bar{z}' \;
e^{-\mu\Delta L}\, e^{(\bar{z}-\bar{z}')\cdot z'} \, \psi(u-u')\, \times
\nonumber \\ \times \, e^{-\omega(x,u, \bar{z},z')\Delta L} \, \delta(x - x' -
u\Delta L)\, q_{j-1}(x',x_{0},u',u_{0}, \bar{z}',z_0) \;,
\end{eqnarray}
where,
\begin{equation}
\psi(u-u')=\mathcal{N}\, e^{-\frac{1}{2\kappa}\Delta L\left(\frac{u-u'}{\Delta
L}\right)^{2}} \;,
\end{equation}
is the angular distribution in velocity space, that tends to orient the 
direction $u$ of a given step with the direction $u'$ of the previous one. In
other words, this distribution introduces the stiffness, where larger values of
$\kappa$ correspond to more flexibility. The interaction term is, 
\begin{eqnarray}
\omega(x,u, \bar{z},z')= \phi(x) -ig u_\mu \Lambda_{\mu}^{A}(x) T^{A}_{cd}\,
\bar{z}^{c} {z'}^{d} +
\Phi^{A}(x) T^{A}_{cd}\, \bar{z}^{c} {z'}^{d},
\end{eqnarray}
and the recurrence is subjected to the initial condition,
\begin{equation}
q_{0}(x,x_{0},u, u_{0}, \bar{z},z_0)=\delta(x-x_{0})\delta(u-u_{0})\,
e^{\bar{z}\cdot z_{0}} \;.
\end{equation}
In this case, upon $M$ iterations we get a weight 
\begin{equation}
q (x,x_{0},u, u_{0}, \bar{z},z_0,L) \equiv q_{M} (x,x_{0},u, u_{0}, \bar{z},z_0)
\;,
\end{equation} 
 that when projected on well-defined initial and final colours gives the
discretized version of the path-integral,
\begin{eqnarray}
\lefteqn{Q^{ba}(x,x_0,u, u_{0},L)= \int
[Dx(s)]_{x_0,x,u,u_0,L}\;[Dz(s)]_{(a,b)}} \nonumber \\
&& \times\, e^{ -\int_{0}^{L} ds\, \left[\mu + \frac{1}{2} (\bar{z}_c \dot{z}_c
- \dot{\bar{z}}_c z_c)+ \frac{1}{2\kappa}\, \dot{u}_\mu \dot{u}_\mu 
+\phi(x) -ig\, \dot{x}_\mu \Lambda_{\mu}^{A}T^{A}_{cd}\, 
\bar{z}^{c} z^{d} +\Phi^{A}T^{A}_{cd}\, 
\bar{z}^{c} z^{d}\right]} \;, 
\end{eqnarray}
where $[Dx(s)]_{x_0,x,u,u_0,L}$ represents the integral over paths with fixed
length $L$ such that, $x(0)=x_0$, $x(L)=x$, $u(0)=u_0$, $u(L)=u$.
That is,
\begin{eqnarray}
\lefteqn{Q^{ba}(x,x_0,u, u_{0},L)=  } \nonumber \\
&& =  \frac{1}{{\cal M}}\, \int dz d\bar{z}\;dz_{0} d\bar{z}_{0}\;  e^{- (\bar{z}\cdot z)/2}\,
e^{-(\bar{z}_{0}\cdot z_0)/2}\,z^{b}\bar{z}_{0}^{a}\; q (x,x_{0},u, u_{0},
\bar{z},z_0,L) \;. \nonumber \\
\label{Qab} 
\end{eqnarray}

From the recurrence relation that generates the discretized weight $q$, we will
be able to derive, in section \ref{diff}, the associated diffusion equation. For
this aim, we note here that taking 
$j=M+1$ in eq. \eqref{recesi},  shifting $x\to x + u \Delta L$, and using the
$\delta$-function to integrate over $x'$,
\begin{eqnarray}
\lefteqn{q(x+u \Delta L,x_{0},u,u_{0}, \bar{z},z_0,L+\Delta L) =\int d^{n-1}u'\,
dz'd \bar{z}'} \nonumber  \\ 
&& \times\, e^{-\mu\Delta L}\, \psi(u-u') \, e^{(\bar{z}-\bar{z}')\cdot z'}
\,e^{-\omega(x,u, \bar{z},z')\Delta L}\, q(x,x_0,u',u_{0}, \bar{z}',z_0,L) \;.
\end{eqnarray}
Now, expanding to first order in $\Delta L$, we get ($\partial_\mu=\frac{\partial ~}{\partial x^\mu}$),
\begin{eqnarray}
\label{expdl}
\lefteqn{q + \Delta L\, (\partial_{L}q +u_\mu\partial_\mu\, q ) =} \nonumber \\
&& = \int dz'd \bar{z}' e^{(\bar{z}-\bar{z}')\cdot z'} \int d^{n-1}u'\,
\psi(u-u') (1-(\mu +\omega(x,u, \bar{z},z')) \Delta L) \nonumber \\
&& \times \,\left[ q' - \langle \Delta u \rangle_{i} \nabla_{u}^{i}q' +
(1/2)\langle \Delta u \otimes \Delta u \rangle_{ij}
\nabla_{u}^{i}\nabla_{u}^{j}q' \right] ,
\end{eqnarray}
\begin{equation}
\langle \Delta u\rangle_i = \int d^{n-1}u'\, (u'-u)_{i}\, \psi(u-u') \;,
\end{equation}
\begin{equation}
\langle\Delta u \otimes \Delta u\rangle_{ij} =\mathcal{N}\int d^{n-1}u'\,
(u'-u)_{i}(u'-u)_{j} \, \psi(u-u') \;,
\label{sm}
\end{equation}
with the abbreviation, $q=q(x,x_0,u,u_{0}, \bar{z},z_0,L)$, $q'=q(x,x_0,u,u_{0},
\bar{z}',z_0,L)$.

\section{Diffusion equation with colour} \label{diff}  

In this section, we shall discuss the first and second order moments of the
distribution $\psi(u-u')$, $u,u' \in S^{n-1}$ and obtain the diffusion equation
in $n$ dimensions. 

Initially, we can suppose that the unit vector $u$ is along the element $e_1$ of
a Cartesian orthonormal basis $e_1,\dots,e_n$, in the space $R^n$ of general
(non-normalized) velocity vectors. The symmetry of the problem implies that
$\langle\Delta u\rangle$ must point along $e_{1}$, while in general it points
along the unit vector $u$. Therefore, the operator $\langle \Delta u \rangle_{i}
\nabla_{u}^{i}$ in eq. \eqref{expdl} is the radial derivative in $R^n$, and
applying it on the function $q(x,u,x_{0},u_{0},z,z_0,L)$, that only depends on
the angular variables, gives a vanishing result. 

A similar argument can be applied to the second order moments. Having a
privileged direction $e_1$, and rotational symmetry on the remaining directions
$e_2,\dots,e_n$,
they have the general structure,
\begin{eqnarray}
\langle\Delta u \otimes\Delta u\rangle = c\; e_{1}\otimes 
e_{1} + h\; (e_{2}\otimes 
e_{2}+\dots + e_{n} \otimes e_{n}) \;.
\label{tensor1}
\end{eqnarray}
The precise value of $c$ is irrelevant for the calculation of the diffusion
equation. Because of the symmetry of the distribution, when $u=e_1$ and $u'$ are
both rotated with a transformation that sends $e_1$ into a general $u$, the second order moments are encoded in,
\begin{eqnarray}
\langle\Delta u \otimes\Delta u\rangle = c\; u\otimes u + h\;
(e_{{\phi}_1}\otimes e_{{\phi}_1} +\dots + e_{{\phi}_{n-1}} \otimes
e_{{\phi}_{n-1}} ) \;,
\label{secmom}
\end{eqnarray}
where $e_{{\phi}_1}, \dots, e_{{\phi}_{n-1}}$ are orthogonal directions, tangent
to $S^{n-1}$ at the point $u$. The tangent directions together with $u$ form an
orthogonal spherical basis in $R^n$. Therefore,
\begin{equation} 
 u\otimes u + e_{{\phi}_1}\otimes e_{{\phi}_1} +\dots + e_{{\phi}_{n-1}} \otimes
e_{{\phi}_{n-1}} = I \;,
 \end{equation}
where $I$ is an $n \times n$ identity matrix, and replacing in eq.
\eqref{secmom},  we get,
\begin{eqnarray}
\langle\Delta u \otimes\Delta u\rangle_{ij} = (c - h)\, u_i u_j + h\, 
\delta_{ij} \;,
\end{eqnarray}
\begin{eqnarray}
\langle\Delta u \otimes\Delta u\rangle_{ij} \nabla_{u}^{i}\nabla_{u}^{j}\, q = 
h\, \nabla_{u}^{2}\, q = h\,\hat{L}_{u}^{2}\, q \;.
\end{eqnarray}
Here, we used again that $q$ only depends on the angular variables in $R^n$, so that
$u_i u_j \nabla_{u}^{i}\nabla_{u}^{j}\, q =0$, and 
\begin{eqnarray}
\label{lapu}
\nabla_{u}^{2}q\equiv\frac{1}{r}\frac{\partial^{2}}{\partial
r^{2}}(rq)+\hat{L}_{u}^{2}q=\hat{L}_{u}^{2}q,
\end{eqnarray}
where $r=|u|$ and $\nabla_{u}^{2}$ is the Laplacian, in the space $R^n$ of
non-normalized velocity vectors, while $\hat{L}_{u}^2$ is the Laplacian on the
$S^{n-1}$ sphere. 

Then, all we need now is computing $h$, and this can be done by taking $u= e_{1}$,
and computing (cf. eq. \eqref{tensor1}), 
\begin{equation}
h = \langle\Delta u \otimes\Delta u\rangle_{22}  \;.
\end{equation}
In spherical coordinates, a general unit vector $u'\in S^{n-1}$ is,
\begin{eqnarray}
\lefteqn{u'=\cos\phi_{1}\, e_{1}+\sin\phi_{1}\cos\phi_{2} \,e_{2}+
\sin\phi_{1}\sin\phi_{2}\cos\phi_{3}\, e_{3} }\nonumber \\
&&+\sin\phi_{1}\cdots\sin\phi_{n-2}\cos\phi_{n-1}\, e_{n-1}+
\sin\phi_{1}\sin\phi_{2}\cdots\sin\phi_{n-2}\sin\phi_{n-1}\, e_{n} \;, \nonumber
\\
\end{eqnarray}
and the integration over $u'$ is done as follows,
\begin{eqnarray}
\int d^{n-1}u'\, f(u')= \prod_{j=1}^{n-2}\int_{0}^{\pi}d\phi_{j}\, (\sin
\phi_{j})^{n-j-1} \int_{0}^{2\pi}d\phi_{n-1}\, f(u')\;.
\end{eqnarray}
For $f\equiv 1$, the integral gives the solid angle $\Omega_n$,
subtended by $S^{n}$,
\begin{eqnarray}
\label{angsol}
\Omega_{n}=\frac{2\pi^{\frac{n}{2}}}{\Gamma \left(\frac{n}{2}\right)} \;.
\end{eqnarray}
In the case where $f(u')$ only depends on $\phi_1$, we have,
\begin{eqnarray}
\int d^{n-1}u'\, f(u')= \Omega_{n-2} \int_{0}^{\pi}d\phi_{1}\sin^{n-2}\phi_{1}\,
f(u') \;.
\end{eqnarray}
and when it only depends on $\phi_1$ and $\phi_2$,
\begin{eqnarray}
\int d^{n-1}u'\, f(u')= \Omega_{n-3} \int_{0}^{\pi}d\phi_1\, (\sin
\phi_{1})^{n-2} \int_{0}^{\pi}d\phi_2\, (\sin \phi_{2})^{n-3} \, f(u')\;.
\end{eqnarray} 
In the latter equation, we have supposed $n\geq 4$. 
Then, using that the angular distribution depends on
$(u'-u)^{2}=2(1-\cos\phi_{1})$,
and $\Delta u_2\, \Delta u_2 = \sin^2\phi_{1}\cos^2\phi_{2} $, we obtain,
\begin{eqnarray}
\mathcal{N}=\frac{(\frac{a}{2})^{\frac{n-2}{2}}
e^{a}\,}{2\pi^{\frac{n}{2}}\Omega_{n-2}\,I_{\frac{n-2}{2}}(a)} \makebox[.5in]{,}
 h= \frac{I_{\frac{n}{2}}(a)\,\sigma }{a\, I_{\frac{n-2}{2}}(a)}\, 
\makebox[.5in]{,} a = \frac{1}{\kappa\Delta L} \;,
\end{eqnarray}
\begin{eqnarray}
\sigma = \frac{\sqrt{\pi}}{2^{n-3}} \frac{ \Gamma
\left(\frac{n-2}{2}\right)\Gamma\left(\frac{n+1}{2}\right)}{\left[
\Gamma\left(\frac{n-1}{2}\right)\right]^2\Gamma\left(\frac{n-3}{2}\right)}
\left\{\frac{4\,
\Gamma\left(n-3\right)}{\Gamma\left(\frac{n-3}{2}\right)}-\frac{
\Gamma\left(n-1\right)}{\Gamma\left(\frac{n+1}{2}\right)}\right\}\;,
\end{eqnarray}
where $I_{\nu}(x)$ is the modified Bessel function.
For completness, we quote the result of the separate calculation for $n=3$,
\begin{equation}
\mathcal{N}=\frac{a}{2\pi(1-e^{-2a})} \makebox[.5in]{,} h=\frac{1}{a}\,
\frac{e^{a}+e^{-a}}{e^{a}-e^{-a}}-\frac{1}{a^{2}} \;.
\end{equation}

Taking the $a\to\infty$ limit, which corresponds to the continuum limit 
$\Delta L\to 0$, together with a finite $\kappa$ (semiflexible limit), the
dominant result is, 
\begin{equation} 
h= \sigma/a =  \kappa \sigma\,\Delta L \;,
\end{equation}  
where $\sigma=2/\pi$, in the interesting dimension $n=4$, while
$\sigma=1$, for $n=3$.

Applying the previous results and the property \eqref{deltacoe} to the
diffusion equation \eqref{expdl}, in the continuum limit, 
\begin{eqnarray}
\lefteqn{\partial_{L}q+u_\mu\partial_\mu\, q = - (\mu+\phi(x))\,q + \frac{\kappa
\sigma}{2}
\hat{L}_{u}^{2}\, q } \nonumber \\
&&   +  \bar{z}^{c}\int dz'd \bar{z}' e^{(\bar{z}-\bar{z}')\cdot z'} (ig
u_{\mu}\Lambda_{\mu}^{A}-
\Phi^{A})\, T_{cd}^{A} \,{z'}^{d}\, q(x,x_0,u,u_0,\bar{z}',z_0,L) \;.\nonumber \\
\end{eqnarray}
Now, noting that,
\[
e^{(\bar{z}-\bar{z}')\cdot z'} {z'}^{d}=\frac{\partial}{\partial
\bar{z}^{d}}e^{(\bar{z}-\bar{z}')\cdot z'} \;,
\] 
\begin{eqnarray}
\int dz'd \bar{z}' e^{(\bar{z}-\bar{z}')\cdot z'} {z'}^{d}\,
q(x,x_0,u,u_0,\bar{z}',z_0,L) = \frac{\partial}{\partial
\bar{z}^{d}}\;q(x,x_0,u,u_0, \bar{z},z_0,L) \;,
\end{eqnarray}
we get the diffusion equation,
\begin{equation}
\partial_{L}q = \left[-\mu-\phi(x) + \frac{\kappa
\sigma}{2}\hat{L}^{2}_{u}-u_{\mu}\partial_{\mu}+ (ig
u_{\mu}\Lambda_{\mu}^{A}-
\Phi^{A})\,T^{A}_{cd}\, \bar{z}^{c}\frac{\partial}{\partial \bar{z}^{d}} \right]
q,
\label{diffz}
\end{equation} 
with the initial condition,
\begin{eqnarray}
q(x,x_0,u,u_0, \bar{z},z_0,0)=\delta(x-x_{0})\delta(u-u_{0})\, 
e^{\bar{z} \cdot z_{0}} \;.
\label{inz}
\end{eqnarray}

Finally, if we expand $q (x,x_{0},u, u_{0}, \bar{z},z_0,L)$ in
powers of the $\bar{z}$ and $z_0$ components, then only linear terms, for each
type of variable, contribute to $Q^{ba}(x,x_0,u, u_{0},L)$ in eq. \eqref{Qab}.
In addition, the last term in eq. \eqref{diffz} does not mix different orders in
$\bar{z}$. That is, integrating both sides of the equation with, 
\begin{eqnarray} 
\frac{1}{{\cal M}}\, \int dz d\bar{z}\;dz_{0} d\bar{z}_{0}\;  e^{- (\bar{z}\cdot z)/2}\,
e^{-(\bar{z}_{0}\cdot z_0)/2}\,z^{b}\bar{z}_{0}^{a}\; \times \;,
\label{intt}
\end{eqnarray}
is equivalent to replacing, 
\begin{equation}
\label{qexpz}
q (x,x_{0},u, u_{0}, \bar{z},z_0,L) \to  \bar{z}^{c}z_{0}^{d}\, Q^{cd}(x,x_0,u,
u_{0},L) \;,
\end{equation}
and then equating the coefficients. This leads to the diffusion equations, 
\begin{equation}
\label{eqdifdepcor}
\left[(\partial_L - (\kappa \sigma/2)\, \hat{L}^{2}_{u}+(\mu +\phi)\,
\delta_{ac} + 
\Phi^{A}T^{A}_{ac}+ u\cdot D_{ac}\right] Q^{cd}(x,x_0,u, u_{0},L)=0 \;,
\end{equation}
which contains the non Abelian covariant derivative, in the given
representation, naturally arising from the calculation,
\begin{equation} 
u\cdot D_{ac}=u_\mu D_\mu^{ac} \makebox[.5in]{,}
D_{\mu}^{ac}=\delta^{ac}\, \partial_{\mu} - ig\Lambda_{\mu}^{A}T^{A}_{ac} \;.
\end{equation}
Similarly, the corresponding initial condition can be obtained by integrating
both sides of eq. \eqref{inz} with \eqref{intt}, or by identifying the appropriate
coefficient in the expansion of $e^{\bar{z} \cdot
z_{0}}$, namely,
\begin{eqnarray}
Q^{cd}(x,x_0,u,u_0,0)= \delta^{cd}\,\delta(x-x_{0})\, \delta(u-u_{0})
\;.
\label{iniab}
\end{eqnarray}

In matrix notation, the coupled diffusion equations for the $Q^{cd}$\,'s can be simply represented as,
\begin{equation}
\label{eqdifdepcor1}
\left[(\partial_L - (\kappa \sigma/2)\, \hat{L}^{2}_{u}+(\mu +\phi )\,
1 + 
\Phi^{A}T^{A} + u\cdot D\right] Q(x,x_0,u, u_{0},L)=0 \;,
\end{equation}
with the initial condition,
\begin{eqnarray}
Q(x,x_0,u,u_0,0)= \delta(x-x_{0})\, \delta(u-u_{0})\, 1 \;,
\label{iniab1}
\end{eqnarray}
where $Q|^{cd} =Q^{cd}$, $1$ is a $\mathscr{D} \times \mathscr{D}$ identity matrix, and
\begin{equation}
D_\mu = 1\, \partial_{\mu} - ig\Lambda_{\mu}^{A}T^{A} \;.
\end{equation} 

\section{Large distance effective field description} 

Now, we can follow similar steps to those given in \cite{ref23}, taking care of
the new dimensionality, the inclusion of colour, as well as the different
boundary conditions we are interested in. In the case of center vortices in $3D$
\cite{ref23}, we looked for the end-to-end probability $Q(x,x_0)$ for an open
one-dimensional object with fixed endpoints and any {\it independent} initial
and final tangent directions. In particular, the equation and the initial
condition describing a diffusion process in $x$ and $u$-space was treated under
this assumption. Then, by joining the initial and final center vortex endpoints in groups of $N$ (for $SU(N)$), to form and instanton/anti-instanton pair,
the weight for a closed chain was obtained as $Q^N(x,x_0)$.

On the other hand, in eq. \eqref{zdefq} we are interested in describing a smooth closed monopole worldline that contains the point $x$, so we need information about the quantity $Q^{ba}(x,x_0,L)$ in eq. \eqref{Qprin}, and then set $b=a$, $x=x_0$ to obtain the desired worldline weight. 
In the language developed in the previous sections, $Q^{cd}(x,x_0,L)$ is to be
identified with the colour components of the matrix,
\begin{eqnarray}
Q(x,x_0,L) =\int d^{n-1}u\;  Q(x,x_0,u,u,L) \;,
\label{lan}
\end{eqnarray}
so that the problem is how to obtain a diffusion equation in $x$-space for this
reduced weight, 
knowing the integrand satisfies eq. \eqref{eqdifdepcor1} and the initial
condition \eqref{iniab1}. 

Initially, we can expand the $u$-dependence of $Q(x,x_0,u, u_{0},L)$ in
eigenfunctions $Y_{lj}(u)$ of the Laplacian operator on $S^{n-1}$, where,
\begin{equation}
\hat{L}^{2}_{u}\, Y_{lj}(u)=  l(l+n-2)\, Y_{lj}(u)\;,
\end{equation}
and $j$ represents the other quantum numbers. In other words, we can expand the
weight in different $l$-sectors,
\begin{equation}
Q(x,x_0,u,u_0,L)=\sum_{l=0} {\cal Q}_{l} (x,x_0,u,u_0,L) \;,
\label{l1}
\end{equation}
where,
\begin{equation}
 {\cal Q}_{l} (x,x_0,u,u_0,L) = \sum_{j=1}^{N(l,n)}  {\cal Q}_{lj}
(x,x_0,u_0,L)\, Y_{lj}(u)  \;,
\label{l2}
\end{equation}
and $N(l,n)$ is the number of linearly independent spherical harmonics with
angular momentum $l$, given by $N(0,n)=1$, and 
\begin{equation}  
N(l,n)=\frac{2\,l+n-2}{l}\, \binom {l+n-3} {l-1} \;,
\end{equation}
for $l\geq 1$. 
With regard to the initial condition \eqref{iniab1}, it is translated to,
\begin{eqnarray}
\sum_{l,j} {\cal Q}_{lj} (x,x_0,u_0,0)\, Y_{lj}(u) =
\delta(x-x_{0})\, \delta(u-u_{0})\, 1
\;,
\end{eqnarray} 
then, using the completeness relation,
\begin{eqnarray}
\delta (u-u_0)&=&\sum_{l,j}\,
[Y_{l,j}(\phi_{1}^0,\phi_{2}^0,\cdots,\phi^0_{n-1})]^{\,\ast} \,
Y_{l,j}(\phi_{1},\phi_{2},\cdots,\phi_{n-1}) \nonumber \\
&=& \sum_{l,j}\, [Y_{l,j}(u_0)]^{\,\ast} \, Y_{l,j}(u)\;,
\end{eqnarray}
it amounts to,
\begin{equation}
{\cal Q}_{lj} (x,x_0,u_0,0) = \delta(x-x_{0})\,
[Y_{l,j}(u_0)]^{\,\ast}\, 1 \;.
\end{equation}
In particular, from eq. \eqref{l2},
\begin{equation}
 {\cal Q}_{0} (x,x_0,u,u_0,0) = \delta(x-x_{0})\,
[Y_{01}(u_0)]^{\,\ast}\, Y_{01}(u)\, 1 =
\Omega^{-1}_{n-1}\, \delta(x-x_{0})\, 1\;.
\label{l3}
\end{equation}

Because of the $u$-factor, when $u\cdot D$ acts on $Q$ in eq. 
\eqref{eqdifdepcor1}, it mixes the different $l$-sectors, and the result can be
again reorganized into $l$-sectors,
\begin{equation}
(u\cdot D)\, Q = \sum_{l=0} (u\cdot D)\, {\cal Q}_{l} = \sum_{l=0}
{\cal R}_{l}\;,
\label{rdef}
\end{equation}
Then, the $l$-components satisfy,
\begin{equation}
\label{eqdifxx0l}
h_l\, {\cal Q}_{l} +  {\cal R}_{l} = -\partial_{L}{\cal
Q}_{l} \;,
\end{equation}
where $h_l$ is a matrix with elements,
\begin{equation}
h^{cd}_l=[ \phi +\mu + l(l+n-2)\, \kappa \sigma/2]\, \delta_{cd}
+\Phi^{A}T^{A}_{cd} \;. 
\end{equation}
For instance,
\begin{equation}
h_0\, {\cal Q}_{0} +  {\cal R}_{0} = -\partial_{L}{\cal
Q}_{0}
\makebox[.5in]{,}
h_1\, {\cal Q}_{1} +  {\cal R}_{1} = -\partial_{L}{\cal
Q}_{1} \;, 
\label{r1}
\end{equation}
\begin{equation}
h_0=[ \phi +\mu ]\, 1 +\Phi^{A} T^{A}  \;,
\end{equation}
\begin{equation}
h_1=[ \phi +\mu + (n-1)\, \kappa \sigma/2]\, 1
+\Phi^{A}T^{A} \;.
\end{equation}
From the definition of ${\cal R}_l$ in eq. \eqref{rdef},
using the properties for the addition of angular momenta, and that the
components of $u$ have $l=1$, we see that ${\cal R}_{0}$ depends on ${\cal
Q}_1$, while ${\cal R}_{1}$ depends on ${\cal Q}_0$ and ${\cal Q}_2$,
\begin{equation}
{\cal R}_{0} = \Big[ (u\cdot D)\,  {\cal Q}_{1} \Big]_0 
\makebox[.5in]{,}
{\cal R}_{1} = \Big[(u\cdot D)\, {\cal Q}_{0} +(u\cdot D)\, {\cal
Q}_{2}\Big]_1\;. 
\label{erres} 
\end{equation}
Now, in the semiflexible case, the memory about the initial orientation $u_0$ is
expected to be erased for large $L$. This means that the final $u$-distribution
will be nearly isotropic, and we can approximate it by taking ${\cal
Q}_{l}\approx 0$, for $l\geq 2$. The memory loss is more effective for
larger values of $\kappa$. Then, in this limit, the four equations in \eqref{r1}
and \eqref{erres} become a closed system, as we can approximate,
\begin{equation}
{\cal R}_{1} \approx \Big[(u\cdot D)\, {\cal Q}_{0} \Big]_1
= (u\cdot D)\, {\cal Q}_{0} \;,
\end{equation}
to obtain,
\begin{equation}
 -\partial_{L}{\cal Q}_{0} = h_0\, {\cal Q}_{0} +  \Big[ (u\cdot D)\, {\cal Q}_{1}
\Big]_0  
\makebox[.3in]{,}
-\partial_{L}{\cal Q}_{1} = h_1\, {\cal Q}_{1}  + (u\cdot D)\,{\cal Q}_{0}  \;.
\label{mdif}
\end{equation}
Writing ${\cal Q}_{1}$ from the second equation, 
applying $ u\cdot D$, and then taking the $l=0$ component, 
\begin{eqnarray} 
 \Big[ (u\cdot D)\, {\cal Q}_{1}\Big]_0   =  - \partial_\mu (h^{-1}_1) \,
\Big[u_\mu\, \partial_{L}{\cal Q}_{1}\Big]_0  -h_1^{-1}\, \partial_L 
\Big[ (u\cdot D) \, {\cal Q}_{1} \Big]_0 &&\nonumber \\
 -\left[u_\mu \, u_\nu \right]_0 \left( \partial_\mu (h^{-1}_1) \,  D_\nu {\cal
Q}_{0} + h^{-1}_1 \, D_\mu D_\nu {\cal Q}_{0} \right)\;. && 
\end{eqnarray} 
For large values of $\kappa$,  
\begin{equation}
h_1^{-1}\approx \frac{\alpha}{\kappa}\, \left[1 - \frac{\alpha}{\kappa} \,(\phi
+\mu)\, 1  -
\frac{\alpha}{\kappa}\, \Phi^{A}T^{A}\right] \makebox[.5in]{,}
\partial_{\mu} h_1^{-1} \sim O(1/\kappa^2)\;,
\end{equation}
$\alpha^{-1} =(n-1)\, \sigma/2$. Then, keeping up to first order terms in
$1/\kappa$,
\begin{eqnarray}
 \Big[ u\cdot D\, {\cal Q}_{1}\Big]_0   \approx -\frac{\alpha}{\kappa n}\,
\left(   D_\mu D_\mu {\cal Q}_{0} \right) -\frac{\alpha}{\kappa}\, \partial_L
\Big[u\cdot D \, {\cal Q}_{1} \Big]_0 
 \;, && 
\label{mdifs}
\end{eqnarray} 
where we used,
\begin{equation}
[u_{\mu}\, u_{\nu}]_0= [u_{\mu}\, u_{\nu}- (1/n)\, \delta_{\mu\nu} + 
(1/n)\, \delta_{\mu\nu}]_0 = (1/n)\, \delta_{\mu\nu}\;.
\end{equation}
So that combining the first equation in \eqref{mdif} and eq. \eqref{mdifs},
\begin{equation}
-\left(1+\frac{\alpha}{\kappa}\, h_0 \right)\partial_{L}{\cal Q}_{0}  \approx
h_0\, {\cal Q}_{0} - \frac{\alpha}{\kappa n}\, D_\mu D_\mu {\cal Q}_{0}
+\frac{\alpha}{\kappa}\, 
\partial^2_{L}{\cal Q}_{0}  \;.
\end{equation}
For large $\kappa$, as $(\alpha h_0/\kappa) << 1$, we get,
\begin{equation}
O {\cal Q}_{0}  + \partial_{L}{\cal Q}_{0} +\frac{\alpha}{\kappa }  \, 
\partial^2_{L} {\cal Q}_{0} \approx 0 \;,
\label{Qeq}
\end{equation}
\begin{equation}
O = - \frac{\alpha}{\kappa n}\, D_\mu D_\mu + (\phi +\mu ) \, 1 +\Phi^{A} T^{A}
\;.
\end{equation}
The third term in eq. \eqref{Qeq} is irrelevant. Formally solving the
characteristic equation for the second order differential equation in $L$, the
large $\kappa$ behaviour of a basis of solutions is,
\begin{equation}
{\cal Q}^{(A)}_{0}(L) =  e^{-L O} \, {\cal Q}^{(A)}_{0}(0) \makebox[.5in]{,}
{\cal Q}^{(B)}_{0}(L) = e^{-\frac{\kappa}{\alpha}\, \left(1-\frac{\alpha}{\kappa} \, O\right)L}\, {\cal
Q}^{(B)}_{0}(0) \;.
\end{equation}
In the approximation where $\kappa$ is much larger than any other mass scale, we have,
$e^{-\frac{\kappa}{\alpha}\, \left(1-\frac{\alpha}{\kappa}\, O\right)L}\approx e^{-\frac{\kappa}{\alpha}\, L}$, as the eigenvalues of $O$ are $\kappa$-independent.
Then, option $B$ is suppressed with respect to option $A$, which simply solves, 
\begin{equation}
O {\cal Q}_{0}  + \partial_{L}{\cal Q}_{0} = 0 \;,
\end{equation}
to be considered with the initial condition (cf. eq. \eqref{l3}), 
\begin{equation}
{\cal Q}_0 (x,x_0,u,u_0,0) =\Omega^{-1}_{n-1}\,\delta(x-x_{0})\, 1\;.
\end{equation}

Summarizing, at large distances, the memory about $u_0$ is lost, and we can
approximate ${\cal Q}$ in eq. \eqref{l1} by the $u$ and $u_0$-independent
quantity ${\cal Q}_0$, 
\begin{equation}
{\cal Q} (x,x_0,u,u_0,L) \approx \Omega^{-1}_{n-1}\, \langle x|e^{-L O} 
|x_0\rangle \;.
\end{equation}
Thus, replacing in eq. \eqref{lan}, 
\begin{equation}
{\cal Q} (x,x_0,L) \approx \langle x|e^{-L O}  |x_0\rangle \;,
\end{equation}
and using eq. \eqref{zdefq}, we finally get,
\begin{eqnarray}
Z &=&\int [D\phi][D\Phi]\, e^{-W}\, e^{\,\sum_{a} \int_{\Re^{4}} d^{4}x \;
\langle x|\int_{0}^{\infty}\frac{dL}{L}\, e^{-L O}  |x\rangle|_{aa} } \nonumber
\\
&=&\int [D\phi][D\Phi]\, e^{-W}\, e^{\, - {\rm Tr} \,  \ln O  }
=\int [D\phi][D\Phi]\, e^{-W}\, ({\rm Det }\, O)^{-1}
\;.
\end{eqnarray}
where ``${\rm Tr}$'' takes the trace in the colour index and the functional trace in
$x$-space, while in the second line we have absorbed a constant factor arising
from the $L$-integration into the measure.
That is, we arrived at the representation,
\begin{eqnarray}
&& Z =\int [D\phi][D\Phi]\, e^{-W} \int [D\zeta][D\bar{\zeta}]\; e^{-\int
d^{4}x\, {\cal L}(\zeta,\Lambda,\phi,\Phi) }\;, 
\end{eqnarray}
\begin{equation}
 {\cal L}(\zeta,\Lambda,\phi,\Phi)= \overline{D_\mu^{ab} \zeta_b} \,  D_\mu^{ac}
\zeta_c + m^2 \,\bar{\zeta}_a \zeta_a +
\frac{\kappa n}{\alpha}\,\Phi^{A} \, \bar{\zeta}_a T^{A}_{ab} \zeta_b
+ \frac{\kappa n}{\alpha}\,\phi \,\bar{\zeta}_a \zeta_a  \;,
\end{equation}
where $m^2=\frac{ n}{\alpha}\,\kappa\mu$, and the components $\zeta_a$, $a=1,
\dots, \mathscr{D}$, form a bosonic complex field in the given representation of $SU(N)$. 

Now, in the adjoint representation $ T^{A}_{bc} \to M^{A}_{BC}= -i
f_{ABC}$. Equivalently, we can introduce the adjoint fields $\zeta =\zeta_{A}\, T_{A}$, $\Phi
=\Phi_{A}\,  T_{A}$, and write,
\begin{equation}
 {\cal L}(\zeta,\Lambda,\phi,\Phi)= \langle D_\mu \zeta ,  D_\mu \zeta\rangle +
m^2 \langle \zeta , \zeta \rangle +
\frac{\kappa n}{\alpha} \langle \Phi , [\zeta ,{\zeta}^\dagger]\rangle 
+ \frac{\kappa n}{\alpha}\phi \,\langle \zeta , \zeta \rangle  \;,
\end{equation}
where the adjoint covariant derivative $D_\mu$, with respect to $\Lambda_\mu =\Lambda^A_\mu
 T_{A}$, and the Lie algebra metric have already been defined in eqs.
\eqref{somed} and \eqref{metric}, respectively.

Therefore, taking for example,
\begin{equation}
W = -\frac{1}{\tilde{\lambda}}\int d^{4}x \, 
\langle \Phi , \Phi\rangle -\frac{1}{\tilde{\eta}}\int d^{4}x \, \phi^{2} \;,
\end{equation}
$\tilde{\lambda} = \frac{\alpha^2}{\kappa^2 n^2}\, \lambda$, $\tilde{\eta} =
\frac{\alpha^2}{\kappa^2 n^2} 
\,\eta$, we arrive at the effective field representation of an ensemble of loops
that carry adjoint colours,
\begin{eqnarray}
&& Z =\int [D\zeta][D\zeta^{\dagger}]\; e^{-\int d^{4}x\, {\cal L}_{\rm
eff}(\zeta,\Lambda) }\;, 
\end{eqnarray}
\begin{equation}
{\cal L}_{\rm eff}(\zeta,\Lambda)= \langle D_\mu \zeta ,  D_\mu \zeta\rangle +
m^2 \langle \zeta , \zeta \rangle +
\frac{\lambda}{4}\, \langle \zeta \wedge {\zeta}^\dagger , \zeta \wedge
{\zeta}^\dagger]\rangle 
+ \frac{\eta}{4}\, \langle \zeta , \zeta \rangle^2  \;,
\end{equation}
or in terms of a pair of hermitian adjoint Higgs fields,  
$\zeta =\frac{1}{\sqrt{2}}\, (\psi_1 + i \psi_2) $,
\begin{eqnarray}
\lefteqn{ {\cal L}_{\rm eff}(\zeta,\Lambda)= \frac{1}{2} \langle D_\mu \psi_I ,
D^\mu \psi_I\rangle  } \nonumber \\
&& + \frac{m^2}{2} \langle \psi_I,\psi_I \rangle + \frac{\lambda}{4}\, \langle
\psi_I\wedge \psi_J,\psi_I \wedge \psi_J\rangle  + \frac{\eta}{4}\, \langle
\psi_I,\psi_I \rangle \langle \psi_J,\psi_J \rangle  \;.
\end{eqnarray}
This result includes all the Higgs field terms of the model in \eqref{modelg4},
\eqref{terms} when
restricted to a pair of flavours. Note that the trilinear term in \eqref{terms} 
involves three different hermitian fields. In our calculation, the obtained squared mass $m^2 =\frac{ n}{\alpha}\, \kappa\mu$ combines the effect of the loop tension and
stiffness, while the quartic $\lambda$-term is associated with isovector loop
interactions. It is important to underline the prominent role played throughout
this work by the finite flexibility $\kappa$, which controls the diffusion in
the space of tangent vectors to the loop. This, together with the non Abelian
loop degrees of freedom, leads to a well-defined continuum limit, and an
effective field model with the usual non Abelian kinetic term showing up at
large distances.

\section{Conclusions} 
 
In this work, we presented a detailed analysis showing how to deal with ensembles of interacting loops carrying non Abelian information in an $n$-dimensional spacetime, for a general group representation with dimension $\mathscr{D}$. Among the interactions, we considered an external non Abelian field, as well as scalar and isovector interactions.

By following a sequence of controlled steps, we were able to go from the initial loop ensemble to the associated non Abelian effective field model. Gaining information about this type of relation could give a complementary perspective on the problem of confinement, as seen by the different communities. This path was made possible by combining recent techniques developed for polymers \cite{ref24} together with the path-integral description of a coloured particle \cite{ref26}. They provided a means to obtain a Chapman-Kolmogorov recurrence relation that generates the end-to-end probability for an open one-dimensional interacting object. Iterating this relation, the object grows in $x \in R^n$ space and in the space of tangent vectors $u \in S^{n-1}$,  governed by the tension $\mu$ and the flexibility $\kappa$, respectively. In particular, the latter is a fundamental property, as it must be present in order to get a meaningful continuum limit. 

Extending the problem from $x$-space to $x$ and $u$-space greatly simplifies the obtention of a diffusion equation for the end-to-end probability. This objective would be very hard to achieve only working in $x$-space. In this case, the simplest noninteracting problems with stiffness have been analyzed by computing the moments of the probability distribution \cite{ref29}, a procedure that is out of scope when interactions are present.

The Chapman-Kolmogorov equation also involves an evolution in the space of non Abelian degrees of freedom $z_a$, $a=1,\dots, \mathscr{D}$, that together with $x$ and $u$ participate in the natural coupling between the loop and the non Abelian gauge 
field. While extended diffusion equations have a simple structure, the diffusion equations in $x$-space are complex. This is due to 
the fact that, in general, the $u$-sector has an infinite tower of possible initial and final angular momentum states. However, when 
$\kappa$ is much larger than any other mass scale (high flexibility), we were able to derive simplified equations. In effect, for large string sizes and flexibility, the memory loss about the initial tangent vector permits an approximation based on the lower values of the final angular momenta.

In this manner, after projecting on the initial and final states with well-defined colours $a,b=1,\dots, \mathscr{D}$, we naturally arrived at a non Abelian diffusion equation for the end-to-end probability. This equation involves a non Abelian Laplacian, as well as scalar and isovector field interactions. 
Then, identifying the initial and final variables in the end-to-end probability, we were able to obtain the loop weight and an effective field description for the ensemble of interacting loops, based on a complex field that transforms in the given group representation.  

Our main result finds a natural application in the analysis of effective 
field models for confinement and their underlying ensembles. On the one hand, we have recently proposed a non Abelian (YMH) model, with adjoint Higgs fields, to describe all the possible confining states between coloured test charges. On the other, in $4D$ lattice Yang-Mills theory, looplike monopoles constitute one of the components of the ensemble of magnetic objects that is expected to drive confinement.  

As the monopole charges are given by the Lie algebra roots \cite{GNO}, and the roots are weights of the adjoint representation, the inclusion of non Abelian information in the ensemble of monopoles should be done using the adjoint. Here, we have shown that the effective field description for an ensemble of interacting adjoint loops precisely contains all the terms involving a given pair of flavours in the YMH confining model. This raises the possibility that, unlike their Abelian counterparts,
non Abelian monopoles could provide an alternative/complementary picture to lattice center vortices as sources of $N$-ality. Further investigations about this relationship will be presented elsewhere.

\section*{Acknowledgements}

We are grateful to G. M. Sim\~oes  for useful discussions. The Conselho Nacional de Desenvolvimento Cient\'{\i}fico e Tecnol\'{o}gico
(CNPq-Brazil) and the Funda\c c\~ao de Amparo a Pesquisa do Estado do Rio de
Janeiro (FAPERJ) are acknowledged for financial support.

\end{document}